\documentclass[%
aip,amsmath,amssymb,
reprint,
author-numerical,
hidelinks]{revtex4-1}
\usepackage[utf8]{inputenc}
\usepackage[T1]{fontenc}
\usepackage{amsmath}
\usepackage{amsfonts}
\usepackage{amssymb}
\usepackage[hidelinks]{hyperref}
\usepackage{empheq}
\usepackage{graphicx}
\usepackage{dcolumn}
\usepackage{mathptmx}

\usepackage{tikz}
\usetikzlibrary{shapes}
\usetikzlibrary{plotmarks}

\newcommand{\bu}{\mathbf{u}}
\newcommand{\bE}{\mathbf{E}}
\newcommand{\bB}{\mathbf{B}}
\newcommand{\bF}{\mathbf{F}}

\newcommand{\bR}{\mathbf{R}}

\newcommand{\wn}{\widetilde{n}}

\newcommand{\wGamma}{\widetilde{\Gamma}}
\newcommand{\wrho}{\widetilde{\rho}}

\begin{document}

\title{Partial-Ionization Deconfinement Effect in Magnetized Plasma}
\date{\today}
\author{M. E. Mlodik} 
\email{mmlodik@princeton.edu}
\author{E. J. Kolmes}
\author{I. E. Ochs} 
\author{T. Rubin}
\author{N. J. Fisch}
\affiliation{Department of Astrophysical Sciences, Princeton University, Princeton, New Jersey, USA, 08543 and \\
Princeton Plasma Physics Laboratory, Princeton, New Jersey, USA, 08540}

\begin{abstract}

	In partially ionized plasma, where ions can be in different ionization states, each charge state can be described as a different fluid for the purpose of multi-ion collisional transport. In the case of two charge states, transport pushes plasma toward equilibrium which is found to be a combination of local charge state equilibrium and generalized pinch relations between ion fluids representing different charge states. Combined, these conditions lead to a dramatic deconfinement of ions. This deconfinement happens on the timescale similar but not identical to the multi-ion cross-field transport timescale, as opposed to electron-ion transport timescale in fully ionized plasma. Deconfinement occurs because local charge state equilibration enforces the disparity in diamagnetic drift velocities of ion fluid components, which in turn leads to the cross-field transport due to ion-ion friction.

\end{abstract}

\maketitle 

\section{Introduction.} 

A new deconfinement mechanism is identified in partially ionized and magnetized plasma. Partially ionized plasma is a plasma where some atomic nuclei retain some of their bound electrons. In partially ionized plasma where ions can be in different ionization states, these states can be modeled as fluids of different species for the purpose of describing multi-ion collisional transport. However, these fluids can transform into one another via ionization and recombination which is not the case in fully ionized plasma. As such, transport in partially ionized plasma is different from transport in fully ionized plasma. In fact, we show here that ionization and recombination combined with multi-ion cross-field transport lead to a dramatic increase of ion deconfinement. We also uncover the new physical mechanism behind it.

Significant progress in understanding of transport in partially ionized plasmas has been made in the parameter regime of the mix of single-ionized ions and neutrals. In particular, relevant transport properties were derived in Ref.  \cite{Helander1994,Catto1994,Hazeltine1992}. These results became a basis to study tokamak scrape-off layer, such as in Ref.\cite{Feng2021,Valovi2004,Zhang2020,Casson2014,McDermott2021,Angioni2014,Angioni2021,Angioni2011,Viezzer2012}. More recently, the same case was also studied in Ref.\cite{Meier2012}. 

The case of multiple charge states is less explored in the literature than is the case of singly ionized plasma and neutrals. More often than not, transport in such cases is found by using ad-hoc diffusion coefficients which match experimental observations, such as in Ref.\cite{Shurygin2006,Lee2022,Odstrcil2017}. In addition to multiple-charge-state effects playing a role  in hot fusion-grade plasmas, particularly in high-Z impurity transport, as indicated above,  these effects may also be expected in low temperature plasma devices, particularly, mass separation devices employing partially-ionized magnetized plasma\cite{Lehnert1971,Krishnan1983,Geva1984,Bittencourt1987,Dolgolenko2017,Ochs2017iii,Gueroult2018,Gueroult2014ii,Gueroult2015,Gueroult2018ii,Gueroult2019,Zweben2018,Rax2016}. Therefore, there is a need to provide a first-principle explanation of transport in partially ionized plasma where ions can be in multiple charge states. This paper identifies the key new mechanisms at play and shows how large and unanticipated new effects may occur.

The difficulty of ions with varying charge states to remain in local thermodynamic equilibrium while respecting the momentum conservation in collisions between magnetized ions results in two kinds of mechanisms. First, there is a larger net transport or deconfinement of all ions. Second, the relative local densities of different charge states may differ from those in thermal equilibrium, affecting inferences of plasma parameters such as electron temperature \cite{Gregorian05a,Gregorian05b}. Thus, deviations from local charge state equilibrium may affect a variety of properties of plasma, such as the time-dependent radial distributions of the ion velocities \cite{Foord}, the magnetic field \cite{Mikitchuk2019,Rosenzweig2020,Davara},  the charge-state composition \cite{Gregorian}, the electron temperature  \cite{Gregorian05a}, the ion temperature \cite{Alumot2019}, and the electron densities  \cite{Gregorian05b, Doron2016} in Z-pinch devices. 
The electron density can be determined from the absolute intensities of spectral lines \cite{Gregorian} and from the ionization times in the plasma \cite{Gregorian05b}. 

If ions are restricted to only two charge states, we prove here that, in fact, ion densities are in local charge state equilibrium, signifying no impact of multi-ion transport on spectroscopic inferences. However, the deviation of relative densities from local plasma equilibrium can be anticipated to appear in the plasma once ions can be in three or more charge states. As such, it could be an important extension of the results presented here to estimate the size of the impact of multi-ion transport in partially ionized plasma on the densities of constituents. The case of three or more charge states, however, is out of the scope of this paper; here, we are content to identify the possibly huge deconfinement effect that is already present in the two ion charge state case. 

The paper is organized as follows. In Sec.~II, equilibrium in the two-charge-state case is considered. In Sec.~III, cross-field transport timescale is derived. 
Sec.~IV summarizes our results and discusses potential applications.

\section{Equilibrium Conditions on Ion Density Profiles.}
\label{sec:equilibriumConditions}

In the case of ions being in only two charge states the density profiles of both charge state fluids can be found exactly. The equilibrium density profiles balance the tension between two types of processes that can change the number of ions in a given charge state at a given spot: non-local and local. The non-local process, which is the cross-field transport, is ambipolar to the leading order, i.e. up to $\mathcal{O}(\sqrt{m_e/m_i})$ or $\mathcal{O}(\rho_i/L)^2$, meaning that if ions can be in charge states $Z_a$ and $Z_b$ ($Z_a < Z_b$) and $\Gamma_s = n_s \bu_s$ is the cross-field particle flux of ions in charge state $s$, then $Z_a \Gamma_a + Z_b \Gamma_b = 0$. This comes from the following observation. In magnetized plasma, force $\bF_s$ acting on species $s$ leads to an $\bF_s \times \bB$ drift, particle flux being 
\begin{gather}
\Gamma_s = \frac{n_s \bF_s \times \bB}{Z_s e B^2}.
\end{gather}  
If $\bF_{ab}$ is the rate of momentum transfer from fluid $b$ to fluid $a$, then $\bF_{ba} = - \bF_{ab}$. Consequently, particle flux due to interaction between fluids $s$ and $s'$, which is 
\begin{gather}
\Gamma_{ss'} = \frac{n_s \bF_{ss'} \times \bB}{Z_s e B^2},
\end{gather}
obeys the following relation: $Z_s \Gamma_{ss'} + Z_{s'} \Gamma_{s's} = 0$. When plasma is out of equilibrium, ion-ion friction force is much larger than electron-ion friction force and viscous forces. Therefore, the corresponding particle flux is also larger and to the leading order $\Gamma_a$ is equal to $\Gamma_{ab}$ while $\Gamma_b$ is  equal to $\Gamma_{ba}$. It also implies that when plasma is pushed out of equilibrium due to a change in external forces acting on plasma, there is timescale separation between ion-ion transport timescale and slower electron-ion transport timescale, as described in Ref.\cite{Mlodik2020}.

\begin{figure}
	\includegraphics[width=0.47\textwidth]{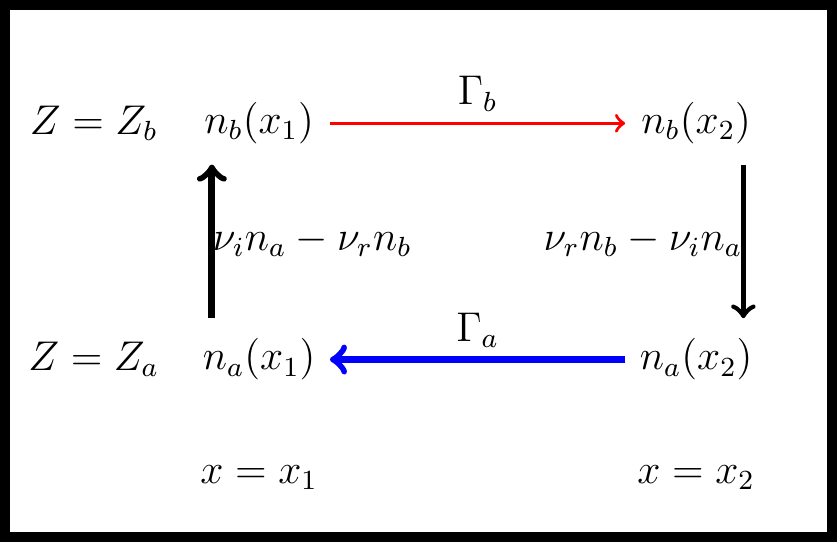}	
	\centering
	\caption{Multi-ion cross-field transport in two-charge-state plasma between locations $x_1$ and $x_2$. Change in density can happen due to ionization or recombination (black lines), transport of ions in charge state $Z_a$ (blue) and $Z_b$ (red). In equilibrium, all the lines should represent the same rate of density change. However, multi-ion cross-field transport obeys ambipolarity condition $Z_a \Gamma_a + Z_b \Gamma_b = 0$. As such, $\Gamma_a = \Gamma_b = 0$ and $\nu_i n_a - \nu_r n_b = 0$ in the equilibrium.}
	\label{fig:conveyorBelt}
\end{figure}

The local processes are ionization and recombination, and they conserve the number of ions, i.e. $n_a + n_b$, locally. If $s$ is the net rate of change of particle density in charge state $b$, then the continuity equation for fluids $a$ and $b$ can be written in equilibrium as
\begin{gather}
\nabla \cdot \Gamma_a = -s;
\end{gather}
\begin{gather}
\nabla \cdot \Gamma_b = s.
\end{gather}
They can be combined to get
\begin{gather}
\nabla \cdot \left( \Gamma_a + \Gamma_b \right) = 0.
\end{gather}
Together with the ambipolarity condition $Z_a \Gamma_a + Z_b \Gamma_b = 0$, this leads to $\Gamma_a = const$. If the boundary condition is no ion flux through the boundary, then particle flux vanishes everywhere, i.e. $\Gamma_a = \Gamma_b = 0$.
Therefore, both non-local and local processes are in the state of dynamic equilibrium when plasma itself is in equilibrium as visualized in Fig.~\ref{fig:conveyorBelt}. In particular, in equilibrium the rate of ionization and recombination are the same. Also in equilibrium there is no cross-field transport due to ion-ion friction. As such, density profiles of charge state fluids satisfy two constraints. The first is generalized pinch relations which are satisfied in the plasma with no net cross-field transport due to ion-ion friction (see Ref.\cite{Taylor1961ii,Spitzer1952,Braginskii1965, Mlodik2020, Mlodik2021,Kolmes2020i, Kolmes2021} for information on generalized pinch relations and Ref.\cite{Kolmes2018} for the derivation). In particular, in the absence of a temperature gradient, they take the form
\begin{gather}
 \big( n_a e^{\Phi_a/T} \big)^{1/Z_a} \propto  \big( n_b e^{\Phi_b/T} \big)^{1/Z_b} ,
 \label{eqn:generalizedPinch}
\end{gather}
where $\Phi_s$ is the potential energy of an ion in charge state $s$. The second constraint is the local charge state equilibrium, which is ensured by the rates of ionization and recombination being equal to each other. Mathematically, local charge state equilibrium can be found in the following way in two limits. In a dilute plasma the corona model
\begin{gather}
\frac{n_a}{n_b} = f(T)
\end{gather}
describes local charge state equilibrium. It can be applied when $10^{12} t_{I}^{-1} < n_e < 10^{16} T_e^{7/2}~cm^{-3}$, where $t_{I} = \left(\alpha_r n_b\right)^{-1} $ is the ionization time, $\alpha_r = 2.7 \times 10^{-13} Z_b^2 T_e^{-1/2} cm^3/s$,  $T_e$ is in the units of $eV$, according to the NRL formulary\cite{Richardson2019}. In a dense plasma, the Saha equilibrium 
\begin{gather}
\frac{n_b n_e^{Z_b-Z_a}}{n_a} = f(T)
\end{gather}
describes local charge state equilibrium. Simultaneously, in this case local charge state equilibrium is equivalent to local thermodynamic equilibrium. According to the NRL formulary\cite{Richardson2019}, electron density is required to be $n_e \gtrsim 7 \times 10^{18} Z_b^7 n^{-17/2} (T_e/E_{\infty}^Z)^{1/2}~cm^{-3}$ in this case if initially an ion in charge state $a$ is in the state $n$ and the ionization energy of that ion is $E_{\infty}^{Z}$. In the case of intermediate density, there are no simplifying assumptions such as the prevalence of two-body or three-body recombination, so codes like FLYCHK\cite{Chung2005} need to be used in order to find the relative abundance of charge states.

These constraints describe an interesting set of density profiles. In thermodynamic equilibrium, when charge state fluids and electron fluid are all in equilibrium,
\begin{gather}
\frac{\nabla n_e}{n_e} = - \frac{\nabla \Phi}{T} \cdot \frac{\langle Z \rangle}{\langle Z(Z+1) \rangle},
\label{eqn:equilibriumElectronDensity}
\end{gather}
where $ \langle ... \rangle $ denotes ion charge state average. In two-charge-state plasma, $\langle X \rangle = \left(X_a n_a + X_b n_b\right)/\left(n_a + n_b\right)$. Eqs.~(\ref{eqn:generalizedPinch}) and (\ref{eqn:equilibriumElectronDensity}) can be combined with the appropriate local charge state equilibrium condition to provide an explicit form of the ion density profiles.

\section{Cross-Field Transport in Two-Charge-State Fluid Model.}
\label{sec:diffusionCoefficientIsothermal}
Suppose that partially ionized plasma is subjected to a change such as an application of potential $\Phi$ and the goal is to find how plasma reacts to that change. In order to isolate the ion deconfinement effect and find out the timescale on which it happens, consider the case when ion-ion transport is much faster than electron-ion transport (as for why this ordering of transport mechanisms is desirable to simplify the model, see Appendix \ref{sec:conveyorBelt}). It is easiest to see the combined effects of collisional cross-field transport and ionization and recombination if plasma possesses the following qualities. First, assume that plasma is an isothermal slab immersed in uniform magnetic field with all the gradients being perpendicular to the slab boundaries. Second, assume that there are no neutral particles in the plasma and ions can be in one of two charge states $a$ and $b$. Third, assume that strength of the magnetic field is such that plasma $\beta$ is low ($\beta \ll 1$) and ion Hall parameter $\Omega_a/\nu_{ab}$ is large ($\Omega_a/\nu_{ab} \gg 1$). Fourth, assume that the effects of ``charge-exchange-like'' collisions, ionization and recombination on transport coefficients are limited to an effective collision frequency such that momentum exchange rate between charge state fluids $a$ and $b$ is $\nu_{ab} m_a n_a \left( \bu_b -\bu_a \right)$ where  $\nu_{ab} = \nu_{ab,elastic} + \nu_{ab,inelastic} + \dot{n_a}/n_a$. Note that in some cases it is impossible to separate elastic and inelastic parts in the collision frequency since ultimately it is a process which obeys the principle of quantum indistinguishability\cite{Krstic1999}. The fourth assumption also implies that the ions which have changed their identity have the fluid velocity of the fluid they were a part of before ionization or recombination. Note, however, that regardless of the details of momentum transfer between species $a$ and $b$, the particle flux due to ion-ion collisions is going to relax to zero in equilibrium because ion-ion transport is ambipolar as long as momentum transfer takes place between the fluids at the same point. There could be a change of shape of equilibrium density profiles by analogy to the effect of thermal force on multi-ion transport. Overall, taking into account the velocity disparity is not going to change the conclusions of Sec.~\ref{sec:equilibriumConditions} and is not going to change the nature of the conclusions of Sec.~\ref{sec:diffusionCoefficientIsothermal}. Fifth, assume that the change of the external potential $\Phi$ is happening fast compared to ion-ion collisional transport timescale but slow compared to faster timescales in the system, and the size of this change is small ($\Delta \Phi \ll T$).

Once all of these assumptions are satisfied, plasma can be treated as a collection of three fluids, two of them representing ions in charge states $a$ and $b$, and the third representing electrons. Note that the timescale of a given cross-field transport mechanism is inversely proportional to the size of particle flux $\Gamma$ generated by it. Therefore, the fastest collisional transport timescale is the ion-ion transport timescale, which is to be derived in this Section. Because the electron-ion friction force and viscous force are much smaller than the ion-ion friction force when plasma is out of equilibrium, the corresponding timescales will be much longer than the ion-ion transport timescale. The change of magnetic field can be neglected due to the low-$\beta$ assumption. Since both charge state fluids are comprised of ions of the same mass, temperature equilibration between them can be expected to happen quickly. Therefore both fluids are assumed to have the same temperature. Also, linearization of fluid equations around the equilibrium found in Sec.~\ref{sec:equilibriumConditions} is possible. 

Despite the fact that the list of necessary assumptions is long, this set of assumptions describes common situations or close approximations to common situations. In particular, in a typical tokamak plasma with tungsten impurities all of the conditions aside from the transport timescale and the assumption that there are only two charge states are satisfied. Moreover, the model can accurately describe collisional transport in helium plasma if the assumptions listed in this Section are satisfied. For example, if parameters similar to an LAPD discharge aside from ion temperature are taken, i.e. $n_e = 10^{12}~cm^{-3}$, $T_i = 7~eV$, $B = 500~G$, then gyroradius of He${}^{+}$ is $\rho_a \approx 1.08~cm$, gyrofrequency of He${}^{+}$ is $\Omega_a \approx 1.2 \cdot 10^6~s^{-1}$, collision frequency of elastic collisions between He${}^{+}$ and He${}^{2+}$ is $\nu_{ab} \approx 1.8 \cdot 10^4~s^{-1}$, plasma beta for ions is $2 \mu_0 \left(n_a + n_b\right) T_i / B^2 \approx 1.2 \cdot 10^{-3}$, charge state abundances, according to corona equilibrium, are around $50\%$. The net ion charge transport timescale, which is derived later in this Section, becomes $\tau_{ci} \approx 23 \cdot \left(\nu_{ab,elastic}/\nu_{ab} \right)~ms$ for a cylindrical plasma with radius $r = 30~cm$, which is comparable to the lifetime of an LAPD discharge, which is on the order of $\sim 10~ms$. 

The following notation is assumed: ions of mass $m$ can be either in charge state $Z_a$ or $Z_b$, density of ions in these states is $n_a$ and $n_b$, respectively. All the equations are linearized around the equilibrium, which is comprised of local charge state equilibrium and generalized pinch relations. 

\begin{figure}
	\includegraphics[width=0.47\textwidth]{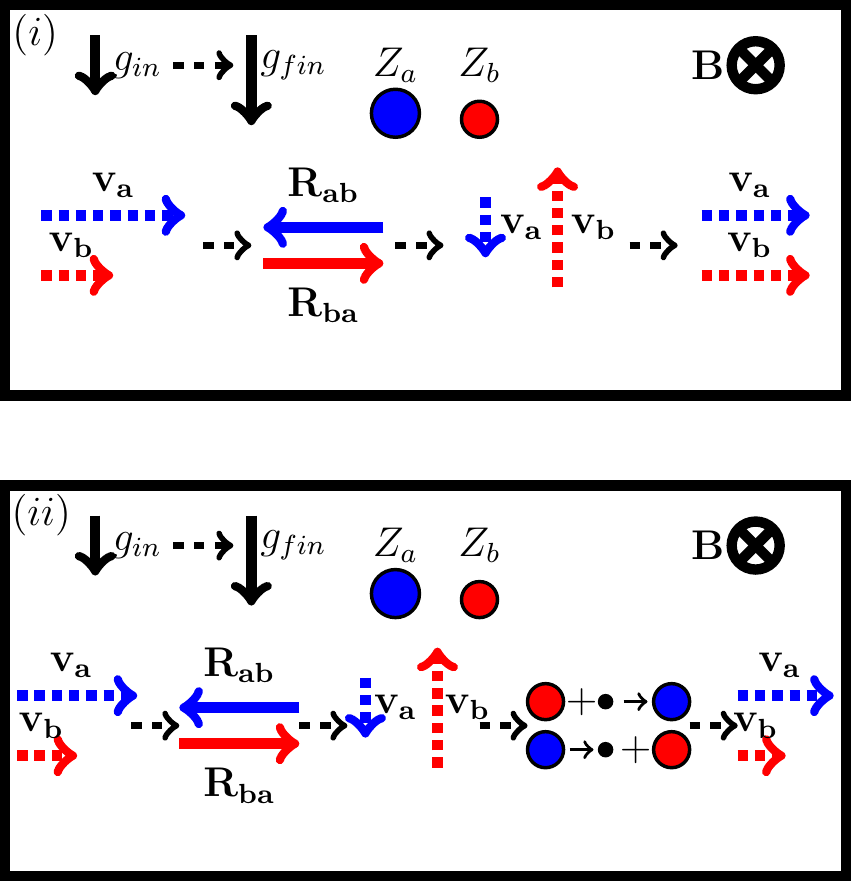}		
	\centering
	\caption{Transport in fully ionized and partially ionized plasma. In fully ionized plasma (i) imbalance in flow velocities leads to ion-ion friction and corresponding cross-field transport. Due to this process ion densities adjust as to relax the difference of flow velocities. In partially ionized plasma (ii), once ions in a charge state move across field lines due to ion-ion friction, they ionize or recombine in order to maintain local charge state equilibrium. Consequently, densities of ion fluids are no longer independent. Because the drift velocities are thus constrained, the plasma must move as a whole.}
	\label{fig:fluidPictureTransport}
\end{figure}

Assume that heat transport across the system is infinitely fast, such that plasma is isothermal. Also assume that density perturbation is small. With these assumptions, the momentum equation for particles of type $s$ is
\begin{gather}
m_s \frac{d_s \bu_s}{dt} = q_s \bE + q_s \bu_s \times \bB - T \frac{\nabla n_s}{n_s} + \frac{\sum_{s'} \bR_{ss'}}{n_s} + \bF_s.
\label{eqn:momentumBasic}
\end{gather}
Here $\bu_s$ is the flow velocity, $d_s/dt = \partial/\partial t + \bu_s \cdot \nabla$, $m_s$ is the mass, $q_s$ is the charge, and $\bR_{ss'} = \nu_{ss'} m_s n_s (\bu_{s'} - \bu_s)$ is the friction force (note that thermal friction is zero under the isothermal plasma assumption), $\bF_s = - \nabla \Phi_s$ is the external force acting on species $s$. It is important to note that the collision frequency $\nu_{ss'}$ in Eq.~(\ref{eqn:momentumBasic}) includes all processes that participate in the momentum transfer between fluids $s$ and $s'$.  Eq.~(\ref{eqn:momentumBasic}) can be rewritten as
\begin{gather}
\bu_s \times \hat{\mathbf{b}} = - \frac{\bE}{B} + \frac{1}{\Omega_s} \frac{d_s \bu_s}{dt} + \frac{T \nabla n_s}{m_s n_s \Omega_s} - \frac{\sum_{s'} \bR_{ss'}}{m_s n_s \Omega_s} - \frac{\bF_s}{m_s \Omega_s}.
\label{eqn:basicMomentumMagnetized}
\end{gather}
The continuity equation for particles in charge state $s$ is
\begin{gather}
\frac{\partial n_s}{\partial t} + \nabla \cdot \Gamma_s = \left(\frac{\partial n_s}{\partial t} \right)_{i/r}.
\label{eqn:continuityBasic}
\end{gather}
Here $\left(\partial n_s/\partial t\right)_{i/r}$ is the change in density of particles in charge state $s$ due to ionization and recombination.
Eqs.~(\ref{eqn:basicMomentumMagnetized}) and (\ref{eqn:continuityBasic}) are linearized around final global equilibrium which is described in Sec.~\ref{sec:equilibriumConditions}. This final equilibrium includes the effect of an external force $\bF_s$ on density profiles. The density of fluid $s$ is split into two components $n_s + \wn_s$ where $n_s$ is density in equilibrium and $\wn_s$ is the difference between the density at the given moment of time and the density in the equilibrium. Under the assumptions made in this section $\wn_s/n_s \ll 1$ everywhere. All the other quantities are split into two parts, equilibrium value and perturbation, in the same way, and the perturbed part is denoted by a tilde. Under the assumption that the external potential is small, i.e. $\Phi/T \ll 1$, density gradients are also small $L \nabla n_s/n_s \ll 1$ so the linearized continuity equations are
\begin{gather}
	\frac{\partial \wn_a}{\partial t} = - \nabla \cdot   \wGamma_{ab} + \nu_r \wn_b - \nu_i \wn_a,
\end{gather}
\begin{gather}
\frac{\partial \wn_b}{\partial t} = - \nabla \cdot   \wGamma_{ba} + \nu_i \wn_a - \nu_r \wn_b.
\end{gather}
Here $\nu_i$ is the effective ionization rate and $\nu_r$ is the effective recombination rate, both of which are calculated as linear terms of $\left(\partial n_s/\partial t\right)_{i/r}$ in density perturbation, taken at equilibrium. $\wGamma_{ss'}$ is particle flux of fluid $s$ due to momentum transfer with fluid $s'$. It is the only component of particle flux that can have non-zero divergence in low-$\beta$ plasma on the ion-ion transport timescale. Given that in the equilibrium $\bu_a = \bu_b$, $\wGamma_{ab}$ is
\begin{gather}
\wGamma_{ab} = \frac{\nu_{ab}}{\Omega_a} n_a \left( \widetilde{\bu}_b - \widetilde{\bu}_a \right) \times \hat{b}.
\end{gather}
Here $\nu_{ab}$ is momentum transfer frequency between fluids $a$ and $b$. Note that $\nu_{ab}$ is higher than it would be if $a$ and $b$ were ions of different species because momentum can be transferred due to ionization or recombination, as well as due to collisions similar to ``charge-exchange'' which are resonant since $a$ and $b$ are ions of same species. $\wGamma_{ba} = - Z_a \wGamma_{ab} / Z_b$ since ion-ion transport is ambipolar.

In the projection of Eq.~(\ref{eqn:basicMomentumMagnetized}) on the direction of gradients the terms $\propto \bF_s$ and $\propto \bE$ are included in the leading order, momentum transfer term $\propto \bR_{ss'}$ is small, polarization drift term $\propto 1/\Omega_s \cdot d_s \bu_s/dt$ is small. Therefore in the first order the only remaining term is diamagnetic drift term. The perturbation of the drift velocity up to the first order is
\begin{gather}
\widetilde{\bu}_s = - \frac{\nabla \wn_s \times \hat{b}}{n_s} \frac{T}{m_s \Omega_s}.
\end{gather}
Linearized continuity equations become
\begin{gather}
	\frac{\partial \wn_a}{\partial t} = \frac{\nu_{ab} T}{m_a \Omega_a^2} \nabla^2 \wn_a - \frac{\nu_{ba} T}{m_b \Omega_a \Omega_b} \nabla^2 \wn_b + \nu_r \wn_b - \nu_i \wn_a,
	\label{eqn:linearizedContinuityA}
\end{gather}
\begin{gather}
\frac{\partial \wn_b}{\partial t} = \frac{\nu_{ba} T}{m_b \Omega_b^2} \nabla^2 \wn_b - \frac{\nu_{ab} T}{m_a \Omega_a \Omega_b} \nabla^2 \wn_a + \nu_i \wn_a - \nu_r \wn_b.
	\label{eqn:linearizedContinuityB}
\end{gather}
Eqs.~(\ref{eqn:linearizedContinuityA}) and (\ref{eqn:linearizedContinuityB}) can be solved by spatial spectral decomposition and then individually for each eigenmode of Laplace operator $\nabla^2$ in the domain of interest. For any particular eigenmode of Laplace operator which has eigenvalue of $-k^2$ (for example, in slab geometry $N$-th mode has eigenvalue $k_N = \pi N/L$), a substitution $\nabla^2 \to -k^2$ 	can be done so the equations for $\wn_a$ and $\wn_b$ become
\begin{gather}
\frac{\partial \wn_a}{\partial t} = - \left( \frac{\nu_{ab} T}{m_a \Omega_a^2} k^2 + \nu_i  \right) \wn_a + \left( \frac{\nu_{ba} T}{m_b \Omega_a \Omega_b} k^2 + \nu_r \right) \wn_b,
\label{eqn:spectralA}
\end{gather}
\begin{gather}
\frac{\partial \wn_b}{\partial t} = - \left( \frac{\nu_{ba} T}{m_b \Omega_b^2} k^2 + \nu_r \right) \wn_b + \left( \frac{\nu_{ab} T}{m_a \Omega_a \Omega_b} k^2 + \nu_i \right) \wn_a.
\label{eqn:spectralB}
\end{gather}
Eqs.~(\ref{eqn:spectralA}) and (\ref{eqn:spectralB}) form a system of two coupled linear ODEs. As such, they have two eigenvalues $\lambda$ which correspond to two exponentially changing solutions ($\propto \exp(\lambda t)$). In particular, if Eqs.~(\ref{eqn:spectralA}) and (\ref{eqn:spectralB}) are rewritten in the form
\begin{gather}
\frac{\partial \wn_a}{\partial t} = - A_{aa} \wn_a +A_{ab} \wn_b,
\end{gather}
\begin{gather}
\frac{\partial \wn_b}{\partial t} =  A_{ba} \wn_a - A_{bb} \wn_b,
\end{gather}
then rates of change of density perturbations are
\begin{gather}
\lambda_{\pm} = - \frac{A_{aa} + A_{bb}}{2} \pm \sqrt{ \left( \frac{A_{aa} + A_{bb}}{2} \right)^2 + A_{ab} A_{ba} - A_{aa} A_{bb} }.
\end{gather}
Momentum conservation implies $m_a n_a \nu_{ab} = m_b n_b \nu_{ba}$. Charge state equilibrium implies $\nu_i n_a = \nu_r n_b$, although this identity is exact when $\nu_i$ and $\nu_r$ are actual ionization and recombination rate, respectively, while we define them as effective rates. Note that this identity still stands in many cases when plasma is close to charge state equilibrium. Given these identities, the following expressions can be obtained:

\begin{gather}
A_{aa} + A_{bb} = \left( \frac{\nu_{ab} T }{m_a \Omega_a^2}  + \frac{\nu_{ba} T }{m_b \Omega_b^2} \right) k^2 + \nu_i + \nu_r,
\end{gather}

\begin{gather}
A_{ab} A_{ba} - A_{aa} A_{bb} = - \frac{k^2 T}{m} \nu_r \nu_{ab} \left( \frac{1}{\Omega_b} - \frac{1}{\Omega_a} \right)^2.
\end{gather}
A useful corollary of these expressions is that $\lambda_{\pm} \leq 0$ in all cases, and $\lambda_{\pm} = 0 $ only if $\nu_i = \nu_r = 0$ (no ionization or recombination). Therefore, the densities of individual plasma components reach equilibrium and there is no instability. Plasma has two modes that decay to equilbrium at the rates $|\lambda_{-}|$ and $|\lambda_{+}|$.

In the limiting case $\nu_i = \nu_r =0$ (when $a$ and $b$ are different species): $\lambda_{+} = 0$,
\begin{gather}
\lambda_{-} = - k^2 \left( \frac{\nu_{ab} T }{m_a \Omega_a^2}  + \frac{\nu_{ba} T }{m_b \Omega_b^2}  \right),
\label{eqn:timescaleFullyIonized}
\end{gather}
which coincides with Ref.\cite{Mlodik2021}. $\lambda_{-}$ has the meaning of the inverse of multi-ion collisional transport timescale in this case.

Another limiting case is a sufficiently large system, such that 
\begin{gather}
\left( \frac{\nu_{ab} T }{m_a \Omega_a^2}  + \frac{\nu_{ba} T }{m_b \Omega_b^2} \right) k^2 \ll \nu_i + \nu_r.
\label{eqn:largeSystemCondition}
\end{gather}
Physically that means that ionization and recombination happen much faster than collisional cross-field transport. Equivalently, this is the case of the infinite, homogeneous plasma ($k \to 0$) where transport is absent. Then
\begin{gather}
\lambda_{-} \approx - \left( \nu_i + \nu_r \right),
\end{gather}
\begin{gather}
\lambda_{+} \approx - \frac{k^2 T}{m} \frac{\nu_r}{\nu_i + \nu_r} \nu_{ab} \left( \frac{1}{\Omega_b} - \frac{1}{\Omega_a} \right)^2.
\end{gather}
In this case, both eigenvalues correspond to processes with a clear physical meaning. To see that, note that if Eq.~(\ref{eqn:largeSystemCondition}) is satisfied and $\lambda = \lambda_{-}$ is used in Eqs.~(\ref{eqn:spectralA}) and (\ref{eqn:spectralB}), then 
\begin{gather}
\frac{\wn_a + \wn_b}{\wn_a} = \frac{\lambda_{-} + A_{aa} + A_{ab}}{A_{ab}} = \mathcal{O} \left( \frac{k^2 T}{m_a \Omega_a^2} \frac{\nu_{ab}}{\nu_i + \nu_r} \right).
\end{gather}
It can be seen that $\lambda_{-}$ represents local charge state equilibration, while $\lambda_{+}$ represents global equilibration associated with cross-field transport. Since $|\lambda_{-}| \gg | \lambda_{+} |$, plasma quickly approaches local charge state equilibrium everywhere, and then adjusts total densities in order to reach global equilibrium. Another corollary of this statement is that spectroscopic inferences from the ratio of densities remain the same even if multi-ion transport is included as long as ions can be only in two charge states and plasma obeys condition set in  Eq.~(\ref{eqn:largeSystemCondition}).

Ion charge transport across magnetic field lines is characterized by the change in total ion charge $\wrho_{ci} = Z_a \wn_a + Z_b \wn_b$ in the slower-varying mode. To see the size of ion charge transport, consider $\lambda = \lambda_{+}$ so
\begin{gather}
\frac{Z_a \wn_a + Z_b \wn_b}{\wn_a} = \frac{Z_a \lambda_{+} + Z_a A_{aa} + Z_b A_{ab}}{A_{ab}}.
\end{gather}
In the large system where Eq.~(\ref{eqn:largeSystemCondition}) is satisfied,
\begin{gather}
\frac{Z_a \wn_a + Z_b \wn_b}{\wn_a} \approx Z_a \frac{\nu_i}{\nu_r} + Z_b.
\end{gather}

This highlights a major difference in the nature of transport in partially ionized plasma. Ion charge density is no longer conserved in such plasma even if cross-field transport happens due to ion-ion friction. This ion deconfinement happens on the timescale
\begin{gather}
\tau_{ci} = \frac{\wrho_{ci}}{\partial \wrho_{ci} / \partial t} = |\lambda_{+}|^{-1}.
\end{gather} Alternatively it can be written as
\begin{align}
 \tau_{ci} = \frac{e^2 B^2}{k^2 T m} \frac{Z_a^2 Z_b^2}{\left(Z_b - Z_a\right)^2} \frac{n_a + n_b}{n_a \nu_{ab}} \left[1 + \mathcal{O} \left(\frac{k^2 T}{m_a \Omega_a^2} \frac{\nu_{ab}}{\nu_i + \nu_r}\right)\right].
\end{align}

Net ion charge moves across field lines dramatically  ($\mathcal{O}(\sqrt{m_i/m_e})$) faster in partially ionized plasma compared to fully ionized plasma. To see it, note that a lower bound on the electron-ion transport timescale in fully ionized plasma is $\tau_{ie,tr} = \left(\tau_{ae,tr}^{-1} + \tau_{be,tr}^{-1}\right)^{-1}$, where $\tau_{ae,tr}$ and $\tau_{be,tr}$ can be found, following Ref.\cite{Mlodik2021}, as RHS of Eq.~(\ref{eqn:timescaleFullyIonized}) with substitutions $b \to e$ and $a \to e$, respectively. After some simplifications using the assumptions $m_a = m_b$ and $m_e/m_a \ll 1$, 
\begin{gather}
\frac{\tau_{ci}}{\tau_{ie,tr}} = \left( \frac{m_a}{2 m_e} \right)^{1/2} \frac{\nu_{ab}}{\nu_{ab,elastic}} \frac{n_a n_b \left(Z_b - Z_a\right)^2}{\left(2 n_e + n_a Z_a^2 + n_b Z_b^2 \right) \left(n_a + n_b\right)} . 
\label{eqn:deconfinementSize}
\end{gather}

Note that Eq.~(\ref{eqn:deconfinementSize}) describes the ratio of net ion charge transport timescales in partially ionized plasma and fully ionized plasma. Unlike in fully ionized plasma where multi-ion transport leads to ion stratification while conserving local ion charge density\cite{Kolmes2018,Mlodik2020}, in partially ionized plasma multi-ion transport leads to plasma moving across magnetic field lines as a whole.
Moreover, this result relies only on the existence of an ion density perturbation in partially ionized plasma. As such, ion deconfinement is going to happen whenever the ion density is out of equilibrium. For example, if plasma consists of background ion species and an impurity species which is present in multiple charge states, then, if the background ion species moves across field lines, the impurity is going to move across magnetic field lines due to the interplay between ionization and collisions between ions in different charge states.

\section{Discussion.} 
\label{sec:Discussion}
Partial-ionization deconfinement effect in magnetized plasma has been identified. Partially ionized plasma has both cross-field transport due to ion-ion friction and ionization and recombination. Combined, these processes result in plasma as a whole being deconfined on multi-ion transport timescale as opposed to fully ionized plasma which is deconfined only on electron-ion transport timescale, $\mathcal{O} (\sqrt{m_e/m_i})$ slower. Virtually all plasmas that include high-$Z$ species are not fully ionized but much of the existing analytic transport theory does not include the possibility of transition between charge states. This paper describes the physical phenomena occurring due to these transitions using first-principles.

The main differences in cross-field transport in partially ionized plasma compared to fully ionized plasma are twofold. First, charge state fluids can exchange momentum between each other not only due to the usual Coulomb collisions but also due to charge-exchange collisions and pick-up current. Second, partially ionized plasma has fewer degrees of freedom to relax the imbalance in flow velocities of fluids which comprise the plasma due to local charge equilibrium being enforced by ionization and recombination which results in extra transport. Perhaps the most important difference between transport in partially ionized plasma and fully ionized plasma is that movement of net ion charge in low-$\beta$ plasma becomes possible even on the ion-ion transport timescale, as opposed to fully ionized plasma where it happens on the electron-ion timescale. This can result in dramatic deconfinement whenever there is a change in external force, such as centrifugal force like in plasma mass filters, acting on plasma. While the deconfinement effect identified here may be the most dramatic effect, it is also noteworthy that there is a charge transport effect that results in net current across field lines, an effect that in and of itself may be quite significant because, in general, so few processes lead to such currents.

In this paper the simplest model is considered in order to isolate ion deconfinement effect. However, there are many ways in which it can be extended. While in two-charge-state case plasma reaches equilibrium when there is local charge state equilibrium and no ion-ion cross-field transport, that would not necessarily be the case if ions can be in three or more charge states. Another extension is adding temperature gradient to the model. Yet another extension could be a problem of background species and partially ionized species. This type of analysis could be important for understanding of high-$Z$ impurity transport.

\acknowledgments{The authors thank R. Gueroult, S. Davidovits, E. Mitra, and J. Waybright for useful conversations. This work was supported by Cornell NNSA 83228-10966 [Prime No. DOE (NNSA) DE-NA0003764] and by DOE DE-SC0016072.}

\section*{Data Availability Statement}

Data sharing is not applicable to this article as no new data were created or analyzed in this study.

\appendix
\section{Electron-Ion Transport Considerations.}
\label{sec:conveyorBelt}
Electron-ion transport and viscous transport are not included in the model considered in this paper in order to isolate the deconfinement effect and not focus on the full dynamics which are complicated due to the following reason. The number of the pathways to change the density of individual charge states limits the problems that can be solved analytically. Consider continuity equation in multispecies plasma:
\begin{align}
\frac{\partial n_s}{\partial t} = - \nabla \cdot \sum_{s'} \Gamma_{ss'} + s_s.
\end{align}
Here $\Gamma_{ss'}$ is particle flux of species $s$ due to collisions with species $s'$, $s_s$ is the source term due to ionization/recombination, different charge states are treated as separate species. Suppose that all ions are of the same chemical element and can be in two charge states $a$ and $b$. If $s$ is the local net rate of ionization and recombination, 
\begin{align}
\frac{\partial n_a}{\partial t} = - \nabla \cdot \left( \Gamma_{aa} + \Gamma_{ab} + \Gamma_{ae} \right) - s.
\end{align}
\begin{align}
\frac{\partial n_b}{\partial t} = - \nabla \cdot \left( \Gamma_{ba} + \Gamma_{bb} + \Gamma_{be} \right) + s.
\end{align}
\begin{align}
\frac{\partial n_e}{\partial t} = - \nabla \cdot \left( \Gamma_{ea} + \Gamma_{eb} + \Gamma_{ee} \right) + s \left(Z_b - Z_a\right).
\end{align}
Clearly, if viscous transport is significant ($\Gamma_{ss}$ is not subdominant), then the problem already becomes complicated. However, since cross-field viscosity is an FLR effect there can be some plasmas where viscous transport is small compared to the frictional transport, e.g. in Ref.\cite{Mlodik2020}. Then
\begin{align}
\frac{\partial n_a}{\partial t} = - \nabla \cdot \left(  \Gamma_{ab} + \Gamma_{ae} \right) - s.
\end{align}
\begin{align}
\frac{\partial n_b}{\partial t} = - \nabla \cdot \left( \Gamma_{ba} + \Gamma_{be} \right) + s.
\end{align}
\begin{align}
\frac{\partial n_e}{\partial t} = - \nabla \cdot \left( \Gamma_{ea} + \Gamma_{eb}  \right) + s \left(Z_a - Z_b\right).
\end{align}
However, there is a degree of freedom remaining, since there are three constraints on four variables ($\Gamma_{ab}$, $\Gamma_{ae}$, $\Gamma_{be}$, $s$). Therefore, even in the presence of the electron-ion frictional transport, individual mechanisms are no longer required to be in the detailed balance (i.e. they don't have to vanish individually in the equilibrium). In turn this means that there can be conveyor-belt-type equilibrium: ionization at the one end of the system, recombination at the other end of the system, and collisional cross-field transport to move ions between those ends. As such, if  electron-ion transport is not negligible compared to ion-ion transport then additional physics may arise, thus obscuring the ion deconfinement effect.

\bibliographystyle{apsrev4-1} 
%\bibliography{../../../Master}
%\bibliography{Master}

\providecommand{\noopsort}[1]{}\providecommand{\singleletter}[1]{#1}%

\end{document}